\documentclass[journal]{IEEEtran}

\usepackage{url}
\usepackage{amsmath}
\usepackage{algorithmic}
\usepackage{algorithm}
\usepackage{booktabs}
\usepackage{tikz}
\usepackage{pgfplots}
\usetikzlibrary{arrows}

\definecolor{plotcolor1}{rgb}{0,0,.8}
\definecolor{plotcolor2}{rgb}{.8,0,0}
\definecolor{plotcolor3}{rgb}{0,.6,0}
\definecolor{plotcolor4}{rgb}{0,.8,.8}
\def\linestyleA{solid}
\def\linestyleB{dashed}
\def\linestyleC{dashdotted}
\def\linestyleD{densely dotted}
\newlength{\figurewidth}
\newlength{\subfigheight}
\newlength{\subfigtop}
\pgfplotsset{
compat=1.14,
width=.85\figurewidth, height=0.35\figurewidth,
scale only axis,
every axis plot/.append style={line width=1pt},
every axis plot/.append style={mark size=2pt},
grid style=dashed,
label style={font=\small},
legend style={font=\scriptsize, inner xsep=2pt, inner ysep=1pt, nodes={inner sep=1pt, text depth=1pt}},
major tick length=3pt,
minor tick num=4,
minor tick length=1pt,
scaled ticks=false,
tick label style={font=\scriptsize},
title style={yshift=-7pt},
xmajorgrids,
ymajorgrids,
/pgf/number format/.cd,
set thousands separator={\,},
/tikz/.cd
}

\newcommand{\figref}[1]{Fig.~\ref{fig:#1}}
\newcommand{\algoref}[1]{Algorithm~\ref{algo:#1}}
\newcommand{\tabref}[1]{Table~\ref{tab:#1}}
\newcommand{\secref}[1]{Section~\ref{sec:#1}}

\newcommand{\horizon}{T}
\newcommand{\timeslot}{t}
\newcommand{\slotsize}{\tau}
\newcommand{\tindex}{t'}
\newcommand{\allagents}{\mathcal{N}}
\newcommand{\cost}{J}
\newcommand{\costof}[1]{\cost^{\text{#1}}}
\newcommand{\children}{\mathcal{M}}
\newcommand{\agent}{i}
\newcommand{\child}{j}
\newcommand{\target}{\Theta}
\newcommand{\power}{\mathbf{x}}
\newcommand{\maxpower}{x^\text{max}}
\newcommand{\minpower}{x^\text{min}}

\newcommand{\energy}{\mathbf{e}}
\newcommand{\maxenergy}{e^\text{max}}
\newcommand{\minenergy}{e^\text{min}}
\newcommand{\price}{\mathbf{\rho}}
\newcommand{\error}{\mathbf{\epsilon}}
\newcommand{\errorbound}{\epsilon^{\text{max}}}
\newcommand{\constraint}{\mathcal{C}}
\newcommand{\constraintof}[1]{\constraint^{\text{#1}}}
\newcommand{\topology}{\mathbf{A}}
\newcommand{\congestionSet}{\mathbf{b}}
\newcommand{\congestionThreshold}{\beta}

\newcommand{\storageEfficiency}{\eta}
\newcommand{\leakage}{\lambda}
\newcommand{\PVMarginalCost}{\gamma}
\newcommand{\cop}{\psi}

\newcommand{\rhabstract}{A distributed, hierarchical, market based approach is introduced to solve the economic dispatch problem. The approach requires only a minimal amount of information to be shared between a central market operator and the end-users. Price signals from the market operator are sent down to end-user device agents, which in turn respond with power schedules. Intermediate congestion agents make sure that local power constraints are satisfied and any potential congestion is avoided by adding local pricing differences. Our results show that in 20\% of the evaluated scenarios the solutions are identical to the global optimum when perfect knowledge is available. In the other 80\% the results are not significantly worse, while providing a higher level of scalability and increasing the consumer's privacy.}
\newcommand{\rhtitle}{Collaboratively Optimizing Power Scheduling and Mitigating Congestion using Local Pricing in a Receding Horizon Market}
\newcommand{\rhkeywords}{Smart Grid, Multi-Agent Systems, Market, Self-Organization}

\begin{document}

\title{\rhtitle}

\pdfinfo{
/Title (\rhtitle)
/Author (Cornelis Jan van Leeuwen, Joost Stam, Arun Subramanian, Koen Kok)
/Keywords (\rhkeywords)
/Abstract (\rhabstract)
/Subject (Multi-agent market-based optimization of smart grids)
}

\author{
\parbox{40mm}{\center
Coen~van~Leeuwen\\
TNO\\
Groningen, Netherlands\\
coen.vanleeuwen@tno.nl}
\parbox{50mm}{\center
Joost Stam\\
University of Twente\\
Enschede, Netherlands\\
j.stam-2@student.utwente.nl}
\parbox{40mm}{\center
Arun Subramanian\\
TNO\\
The Hague, Netherlands\\
arun.subramanian@tno.nl}
\parbox{40mm}{\center
Koen Kok\\
TNO\\
Groningen, Netherlands\\
koen.kok@tno.nl}
}

\maketitle

\begin{abstract}
\rhabstract{}
\end{abstract}


\section{Introduction}
\label{sec:lprh:introduction}
Three trends set a challenge for future power grids. Firstly, the transition towards sustainable energy sources leads to more renewable energy, but also to a larger fraction of unpredictable and intermittent production. Secondly, the electrification of various systems such as transport (electric vehicles), heating (heat pumps), and in general an increase of electricity-consuming devices leads to a huge growth of power consumption. And thirdly, the distribution of energy generation leads to a very different pattern in the load of the power transmission grid, than it was designed for.

The control of a vast number of small power units, both consuming and producing, is extremely difficult to do completely top-down, so a centralized control strategy cannot be used~\cite{Kok2016}. At the same time the power infrastructure is aging, and was not built for the emerging pattern of distributed \emph{prosumers}~\cite{Gemine2017}. This is why we need self-organizing control algorithms to schedule the use of electric devices while taking into account constraints of the power distribution infrastructure. This aids \emph{distribution system operators} (DSO) and \emph{transport system operators} (TSO) to maintain power balance and make sure there is no congestion, i.e. the grid capacity is not overloaded.

\begin{table*}[t!]
\begin{center}
\caption{Used notations and parameters in this paper{}.\label{tab:lprh:notation}}
\begin{tabular}{l p{57mm} l l}
\toprule
Symbol & Description & Typical value & Unit \\ \midrule

$\topology$ & topology matrix of child relations & & \\
$\alpha$ & forecast accuracy & $0.9$ & \\
$\congestionSet$ & vector of congestion thresholds & $[2,3,\ldots,3.5] \times 10^4$ & W \\
$\congestionThreshold_\agent$ & congestion threshold of $\agent$ & $3 \times 10^4$ & W \\
$\constraint_\agent$ & constraint function of $\agent$ & & \\
$\constraintof{Z}_\agent$ & constraint function of $\agent$ of type Z$^*$ & & \\
$\PVMarginalCost_\agent$ & PV operation cost of $\agent$ & $0.2$ & \\
$\energy_\agent$ & $1 \times \horizon$ vector of energy of $\agent$ & $[0.0, 1.1, \ldots, -0.3] \times 10^3$ & Wh \\
$e_{\agent\timeslot}$ & energy of $\agent$ at $\timeslot$ & $200$ & Wh \\
$\minenergy_\agent$ & minimum energy of $\agent$ & $0.0$ & Wh \\
$\maxenergy_\agent$ & maximum energy of $\agent$ & $2 \times 10^3$ & Wh \\
$\error$ & $1 \times \horizon$ vector of errors & $[0,11,\ldots,-225]$ & W \\
$\errorbound$ & upper bound on the error & $10^{-3}$ & W \\
$\storageEfficiency_\agent$ & storage efficiency of $\agent$ & $0.9$ & \\
$\target$ & target power of the cluster & $[1.9, -1, \ldots, 2.6] \times 10^4$ & W \\
$\agent$ & an agent index & $0,1,\ldots,n$ & \\
$\cost_\agent$ & cost function of $\agent$ & & \\
$\costof{Z}_\agent$ & cost of $\agent$ of type Z$^*$ & & \\
$\leakage_\agent$ & storage leakage of $\agent$ & $360$ & W \\
$\children_\agent$ & set of children of $\agent$ & & \\
$\allagents$ & set of all agents & & \\
$\price$ & $1 \times \horizon$ vector of prices & $[0.5, 0.3, \ldots, 0.8]$ & \\
$\horizon$ & time horizon & $24$ &\\
$\timeslot$ & time slot index & $0,1,\ldots,\horizon$ & \\
$\slotsize$ & duration of a time slot & 1 & h \\
$\power_\agent$ & $1 \times \horizon$ vector of scheduled powers of $\agent$ & $[1.1, -0.5, \ldots, 0.4] \times 10^3$ & W \\
$\bar{\power}$ & $1 \times \horizon$ vector of aggregated powers of child nodes & $[24.0, -18.6, \ldots, 8.7] \times 10^3$ & W \\
$\power'_\agent$ & $1 \times \horizon$ vector of expected powers of $\agent$ & $[1.0, -0.4, \ldots, 0.8] \times 10^3$ & W \\
$\hat{\power}$ & $1 \times \horizon$ vector of average historic powers & $[0.8, 0.6, \ldots, -0.5] \times 10^3$ & W \\
$x_{\agent\timeslot}$ & scheduled power of $\agent$ at $\timeslot$ & $-1.7 \times 10^{3}$ & W \\
$\minpower_\agent$ & minimum power for $\agent$ & $-2\times10^3$ & W \\
$\maxpower_\agent$ & maximum power for $\agent$ & $4\times10^3$ & W \\
$\cop_\agent$ & heat pump coefficient of performance (COP) of $\agent$ & $4.0$ & \\
\bottomrule
\end{tabular}
\end{center}

Note that powers are indicated from the grid perspective. That means positive powers indicate power going from the grid to the device, and negative powers are from the device back to the grid.

$^*$Z indicates a type of agent, which can either be MO for market operator, CO for congestion, LOAD for a static load agent, PV for an agent with photovoltaic solar panels, or ST for a storage agent.
\end{table*}

\section{Problem Statement}
The problem at hand is a variation of the economic power dispatch problem~\cite{Ross1980,Abido2003}, where the operation of a set of generators is optimized, such that power is provided to consumers in the most cost-effective manner. Traditionally this problem would include controllable generators (power generation plants), constraining transmission resources (transformers, stations, cables), and end-consumers having a static load. Currently, with distributed energy resources, and demand response---the possibility to control the load of consumers using flexibility of smart devices---the problem changes significantly.

A formal definition of the problem is as follows:

\begin{subequations}
\label{eq:lprh:optim}
\begin{align}
\min_{\power}\quad & \sum_{\agent \in \allagents} {\cost_\agent(\power{}_\agent)}, \label{eq:lprh:mincost} \\
\text{s.t.}\quad & \sum_{\agent \in \allagents} {\power_\agent} = \target, \label{eq:lprh:constraint:target}\\
& \topology\power \leq \congestionSet, \label{eq:lprh:constraint:grid}\\
& \minpower_\agent \leq \power_\agent \leq \maxpower_\agent, & \forall \agent \in \allagents, \label{eq:lprh:constraint:power} \\
& \constraint_\agent(\power_\agent) = 0, & \forall \agent \in \allagents. \label{eq:lprh:constraint:agent}
\end{align}
\end{subequations}

With all parameters defined in \tabref{lprh:notation}, the objective function as defined in \eqref{eq:lprh:optim} is the same as for the traditional economic dispatch, which can be summarized as: to find a set of electrical powers $\power_\agent$ for every device $\agent \in \allagents$ in the grid, that minimizes the sum of all costs $\cost_\agent(\power_\agent)$. In the original dispatch problem defined, $\allagents$ defines only the producers. However, in our problem formulation \eqref{eq:lprh:optim} $\allagents$ denotes the full set of devices in the grid, including \emph{producers and consumers}. This means that contrarily to the traditional economic dispatch problem, not only the generators are controllable, but also the flexible loads of end consumers.

Note that the costs in~\eqref{eq:lprh:mincost} do not necessarily refer to the financial cost of a dispatch. Rather, this generic model only optimizes some \emph{social welfare}, which defines a desirable outcome for all participants involved. Depending on the situation at hand, one could minimize the amount of greenhouse gasses emitted, or maximize the amount of renewable energy used. However, for the rest of this paper{} the costs are defined as the energy losses of the devices. By minimizing the amount of energy losses we attempt to find a dispatch that is both economic and sustainable.

Let us discuss the constraints of \eqref{eq:lprh:optim}, i.e. (\ref{eq:lprh:constraint:target}--\ref{eq:lprh:constraint:agent}). The constraint~\eqref{eq:lprh:constraint:target} states that the sum of all powers in the system has to meet a specific target $\target$. In a balanced (island) grid this target has to be zero for every time slot, but in a connected grid this target is the contracted load with the transmission operator. Put differently, $\target$ is the net power input of the grid to the rest of the world.

Constraint~\eqref{eq:lprh:constraint:grid} defines the limitations of the physical power grid; a topology matrix $A$ specifies which device is connected to which grid components, and $b$ is the rating of those components---the maximum amount of power that it can safely transmit.

Constraints~\eqref{eq:lprh:constraint:power} and~\eqref{eq:lprh:constraint:agent} represent the constraints of the devices in the grid. Specifically, constraint~\eqref{eq:lprh:constraint:power} states that a device cannot produce or consume more power than it physically can, which is represented by lower and upper bound, $\minpower_\agent$ and $\maxpower_\agent$, respectively. However,~\eqref{eq:lprh:constraint:agent} also takes into account another \emph{private} constraint $\constraint$. We refer to this constraint as private, as it only concerns the state of a single device, and its state may concern information that we should not require to share with other parties. E.g. an end-user should not need to share its intent to run his or her washing machine with the rest of the neighborhood. This constraint differs per device, and specifies device limitations such as a battery that cannot hold more than a certain amount of energy, and cannot discharge when already empty. Flexible loads also may have constraints considering the time at which they can turn on or off based on the consumer's settings. We will elaborate on these constraints, as well as the cost functions of the device agents in \secref{lprh:agentbehaviour}.

\subsection{Related Work}
\label{sec:lprh:related_work}
In existing studies, different strategies are used for energy management. Four different main categories are defined in~\cite{Kok2016} based on whether there is a distributed aspect of decision-making and whether there is one- or two-way communication. One of the strategies defined in~\cite{Kok2016} uses \emph{Transactive Control}, where distributed systems decide locally on their device management, using two-way communication in a market-based control scheme. The authors compare this approach with traditional top-down switching, price-reaction and centralized optimization strategies and show that transactive control is capable of using the full flexibility potential of the smart grid devices, while maintaining the end-user privacy. This claim is consistent with earlier studies~\cite{Ygge1999,Akkermans2004} that have shown that using a market-based control in a multi-agent system can provide equally optimal results as a centrally optimized system, under certain conditions.

Multi-agent based methods are already getting attention in the smart grid domain due to many desirable properties: robustness, user-friendliness, attack resistance and scalability~\cite{Pipattanasomporn2009,Colson2013}. The economic dispatch problem is very well suited to be represented using multi-agent systems~\cite{Cai2012}. There are many studies that use a multi-agent based approach to model and solve the problem, such as the two-way message passing agents using consensus algorithms to find an allocation that is optimal~\cite{Hu2018}. Other methods use a completely decentralized method involving reinforcement~\cite{Liu2018}, utility maximization~\cite{Li2011} or model predictive control~\cite{Velasquez2019}.

In~\cite{Warrington2011}, a strategy is proposed to schedule the power consumption and production of generators and loads based on a method called \emph{Negotiated Predictive Dispatch}. In this approach wind and conventional generators, as well as static and flexible loads, are controlled on the transmission level. Agents propose power schedules, which are aggregated by a central market operator, which then updates a price program using gradient descent, meaning the price of congested timeslots increase, and that of underused time slots decrease. In doing so the authors show that they are able to balance power production and consumption, while satisfying power grid constraints. However, this approach focuses on the transmission level, whereas we consider the distribution level power grid to be at a much more imminent risk of congestion. At the transmission level, the scale is much larger than at the distribution level, both geographically and considering the power levels involved. Moreover, the power grid has a very different topology at the transmission level. A major drawback of this approach when applied at the distribution level, is that using a global power price for congestion mitigation is not only unfair for agents in non-congested areas, but it would also be converging to a solution much too slowly. A similar approach to the negotiated dispatch is provided by~\cite{Binetti2014} in which a gradient descent on the pricing is used, followed by a local optimization of agents.

Another approach is given by~\cite{vanderKlauw2016}, in which power programs are proposed by device agents that are able to satisfy cluster constraints to reach a target energy consumption, while also staying within the limits of the grid. In their approach, called \emph{Profile Steering}, agents are only motivated to accomplish the cluster goal, which is to consume or produce energy at a specific target amount, and will propose alternative power programs whenever constraints are violated. Proposals that reduce the constraint violations the most, are then selected as the new candidate programs. In our view, the problem of this approach lies in the lack of motivation for the participant to sacrifice private rewards for running an alternative program. The only objective is the cluster goal, which means that the price of having limited resources is paid by a select set of individuals that offer the most flexibility, or conversely, the profit is reaped by those who are able to maximally make use of remaining capacity.

The authors of~\cite{Fioretto2017} propose an approach based on DCOPs (Distributed Constraint Optimization Problem) to solve the economic dispatch problem, as well as the real-time demand-response balancing problem. Their algorithm is able to find optimal solutions for a system of controllable generators and (predicted) loads, taking into account transmission network constraints. They show that finding the optimal solution using Dynamic DCOPs is possible, but their solution does not scale very well. A relaxed version of the problem, in which soft constraints may be (temporarily) violated, scales better, but still for relatively small time horizons, even considering their implemented solution on a GPU.

We note that related work on the topic of this chapter listed here is not exhaustive. For further detailed background, a comprehensive overview of different mechanisms for solving optimization problems in smart grids, particularly where demand response is involved, we refer to~\cite{Vardakas2014,Parhizi2015,Jordehi2019}.

\section{Self-Organizing Economic Dispatch}
\label{sec:lprh:methods}
Our contribution is a self-organizing method for coordinating the power scheduling of smart grid devices on the distribution level energy grid. This is a multi-agent based heuristic approach for find a solution to the economic dispatch problem as defined in \eqref{eq:lprh:optim}. This multi-agent based approach allows for great scalability, while ensuring privacy and final control of the end-user.

At the distribution grid level, devices are typically consumer devices, such as PV-panels, household batteries, heat pumps, ventilation or air-conditioning units. The controllability, or flexibility of such devices is often limited, and bound by the device limitations and the user preferences. With increasing numbers of such devices, using a strictly ``top-down'' approach is intractable, since the search space grows exponentially with each added device. For this reason we use a hierarchical approach, in which the market operator delegates its task to intermediate ``congestion'' agents which then independently solve subproblems---which is to make sure that the total power throughput at their allocated point does not exceed a certain threshold. Making such subdivisions is justified, since in the power grid, transformers are effectively branches of the topology, and two nodes under different transformers are independent of one another; hence, transformers are the logical point to place congestion agents in the control hierarchy.

Our approach, denoted as Local Pricing Receding Horizon (LP-RH) is based on the economic incentive that end-user devices should have, to provide its owner with a service, in the most affordable way. Put differently, the system will use price differences to stimulate agents to schedule their power consumption or production in a balanced way, such that any grid constraints are satisfied. The overall scheme is simply to gather expected power programs from the connected devices, and then iteratively adjust energy prices to steer the agents into a certain power program. The method is explained in more detail in the following section, and is similar to the Negotiated Price Dispatch proposed by~\cite{Warrington2011}.

\subsection{Local Pricing Receding Horizon}

\newlength{\nodedist}
\setlength{\nodedist}{21pt}
\newlength{\nodedistB}
\setlength{\nodedistB}{27pt}
\newlength{\nodedistC}
\setlength{\nodedistC}{33pt}
\newlength{\nodedistD}
\setlength{\nodedistD}{38pt}

\begin{figure}[t]
\centering
\begin{tikzpicture}[auto,rect/.style={rectangle,draw,fill=white,minimum size=0,inner sep=3pt}]
\node[rect] (1) [fill=green] {$\timeslot_1$};
\node[rect] (2) [right of=1,node distance=\nodedist] {$\timeslot_2$};
\node[rect] (3) [right of=2,node distance=\nodedist] {$\timeslot_3$};
\node[rect] (4) [right of=3,node distance=\nodedistB,draw=white] {\ldots};

\node[rect] (5) [right of=4,node distance=\nodedistC] {$\timeslot_{\horizon-2}$};
\node[rect] (6) [right of=5,node distance=\nodedistC] {$\timeslot_{\horizon-1}$};
\node[rect] (7) [right of=6,node distance=\nodedistB] {$\timeslot_{\horizon}$};

\node[rect] (8) [below of=2,node distance=\nodedist,fill=green] {$\timeslot_2$};
\node[rect] (9) [right of=8,node distance=\nodedist] {$\timeslot_3$};
\node[rect] (10) [right of=9,node distance=\nodedist] {$\timeslot_4$};
\node[rect] (11) [right of=10,,node distance=\nodedistC,draw=white] {\ldots};

\node[rect] (12) [below of=6,node distance=\nodedist] {$\timeslot_{\horizon-1}$};
\node[rect] (13) [right of=12,node distance=\nodedistB] {$\timeslot_{\horizon}$};
\node[rect] (14) [right of=13,node distance=\nodedistB] {$\timeslot_{\horizon+1}$};

\node[rect] (15) [below of=9,node distance=\nodedist,fill=green] {$\timeslot_3$};
\node[rect] (16) [right of=15,node distance=\nodedist] {$\timeslot_4$};
\node[rect] (17) [right of=16,node distance=\nodedist] {$\timeslot_5$};
\node[rect] (18) [right of=17,node distance=\nodedistD,draw=white] {\ldots};

\node[rect] (19) [below of=13,node distance=\nodedist] {$\timeslot_{\horizon}$};
\node[rect] (20) [right of=19,node distance=\nodedistB] {$\timeslot_{\horizon+1}$};
\node[rect] (21) [right of=20,node distance=\nodedistC] {$\timeslot_{\horizon+2}$};

\path
(1) edge (2)
(2) edge (3)
(3) edge (4)
(4) edge (5)
(5) edge (6)
(6) edge (7)

(8) edge (9)
(9) edge (10)
(10) edge (11)
(11) edge (12)
(12) edge (13)
(13) edge (14)

(15) edge (16)
(16) edge (17)
(17) edge (18)
(18) edge (19)
(19) edge (20)
(20) edge (21);

\path[->,>=stealth']
(1) edge[bend right=45] (8)
(8) edge[bend right=45] (15);
\end{tikzpicture}
\caption{The principle of a receding horizon market is that every time step the time horizon for optimization shifts by one. Planned power programs (white) are turned into fixed contracts for the current time step (green), which determines the outcome of the algorithm. This figure is redrawn from~\cite{Warrington2011}.}
\label{fig:lprh:receding-horizon}
\end{figure}
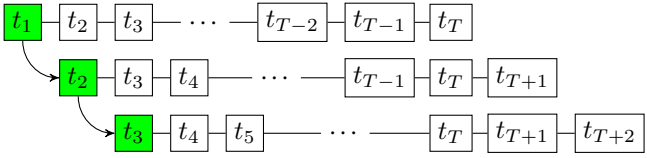

In order to create a power planning taking into account forecasts and/or predicted power programs of clients, we use a Receding Horizon (RH) approach, depicted in \figref{lprh:receding-horizon} which means that at any point we only take a fixed horizon of $\horizon$ program time units (PTU) into account. A PTU is typically 15 minutes or one hour, throughout this paper we will use PTU length of one hour so $\horizon=24$. However, none of the mentioned approaches are limited to this convention. When the algorithm has converged and a power program is found for the next $\horizon$ PTUs that satisfies all constraints, the first PTU becomes the ``current'' situation, and the projected power program becomes a contract. The time horizon now shifts by one, and the entire system starts again.

\begin{algorithm}[t]
\caption{\label{algo:lprh:lprh}LP-RH Algorithm}
\begin{algorithmic}[1]
\item[] $\power_\agent = \text{createPowerprogram}(\agent, \price)$
\IF{$\agent$ is a Device agent}
\STATE $\power_{\agent} = \text{arg min}_{\power} {\constraint_{\agent}(\power,\price)}$ {\color{gray} \COMMENT{Run local optimization}} \label{line:lprh:localoptimization}
\ELSE
\REPEAT
\FORALL{$\child \in \children_\agent$}
\STATE $\power_{\child} = \text{createPowerprogram}(\child, \price)$ \label{line:lprh:createprogram}
\ENDFOR
\STATE $\power_\agent = \sum_{\child \in \children_{\agent}}{\power_\child}$
\STATE $\error = \constraint_{\agent}(\power_{\agent})$ \label{line:lprh:constraint}
\STATE $\price = \text{adjustPrices}(\error, \price)$ \label{line:lprh:adjustprice} {\color{gray} \COMMENT{Using gradient descent}}
\UNTIL{$\error \leq \errorbound$}
\ENDIF
\RETURN $\power_{\agent}$
\end{algorithmic}
\end{algorithm}

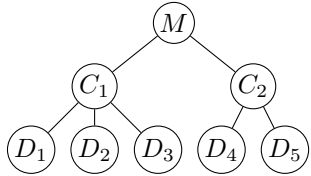
\begin{figure}
\begin{center}
\begin{tikzpicture}[auto,
node/.style={circle,draw,fill=none,minimum size=0, inner sep=1pt},
spacer/.style={circle,draw,fill=white,draw=white}]
\node[node] at (0,0) (m) {$M$};
\node[node] at (-30pt, -24pt) (c1) {$C_1$};
\node[node] at (30pt, -24pt) (c2) {$C_2$};

\node[node] at (-54pt, -48pt) (d1) {$D_1$};
\node[node] at (-30pt, -48pt) (d2) {$D_2$};
\node[node] at (-6pt, -48pt) (d3) {$D_3$};

\node[node] at (18pt, -48pt) (d4) {$D_4$};
\node[node] at (42pt, -48pt) (d5) {$D_5$};
\path
(m) edge (c1) edge (c2)
(c1) edge (d1) edge (d2) edge (d3)
(c2) edge (d4) edge (d5);
\end{tikzpicture}
\caption{Simple example topology of a graph which could be running the LP-RH algorithm. One market operator ($M$) is connected to two congestion agents ($C_\agent$), which are connected to a total of five device agents ($D_\agent$). The device agents can represent PV panels, consumer loads, storage agents, or any other leaf nodes in the power grid. Topologies where congestion agents are connected to more congestion agents are also possible.}
\label{fig:lprh:constraintgraph}
\end{center}
\end{figure}

The Local Pricing Receding Horizon algorithm is shown as pseudocode in \algoref{lprh:lprh}; an example graph on which the algorithm could run is shown in \figref{lprh:constraintgraph}. The algorithm describes how any agent finds a new power program $\power_{\agent}$ to satisfy its local constraint $\constraint_{\agent}$. Every device agent in the grid does this by solving a local optimization problem (line~\ref{line:lprh:localoptimization}), taking into account local constraints as explained in \secref{lprh:agentbehaviour}.

Prices $\price$ are updated by the market operator agent and the congestion agents in order to get a power program $\power$ satisfying the grid constraints. When the market operator or any congestion agent runs the LP-RH algorithm, all $\child \in \children_\agent$, where $\children_\agent$ are the immediate children of agent $\agent$, are requested to propose a power program based on the price $\price$ (line~\ref{line:lprh:createprogram}). Here the function is called recursively until all agents have determined a new power program. Note, that if any of the children are congestion agents, they determine their power program by forwarding the prices to \emph{their} children and returning the sum of all received programs. Many agents will not know exactly how they will function in the future, as contexts might change, a user might behave differently than anticipated, or weather conditions may act up; this is why receding horizon method updates the time iteratively. With the sum of all power programs, the market operator can determine the error $\error$, which is the sum of all local constraint violations $\constraint$ for every $\timeslot \in \horizon$.

In line~\ref{line:lprh:constraint} the constraint $\constraint_{\agent}$ is used to compute the local constraint violation and stored as an error variable $\error$. The value of $\error$ is taken into account to adjust the prices in line~\ref{line:lprh:adjustprice}. In the ``adjustPrices'' function, a gradient descent approach is used, which linearly interpolates the last two errors as a function of the price. We then choose the price at which the error is projected to reach zero. Note that this means we assume the reaction of the devices linearly depends on the prices, which will only hold under very specific circumstances. To overcome this issue, we iteratively repeat the process until $\error \leq \errorbound$ and assume that a non-linear response can be described as a series of linear pieces. The parameter $\errorbound$ then represents an upper bound on the error, which we can use as a convergence criterion.

\subsection{Agent Behavior}
\label{sec:lprh:agentbehaviour}
Agents in the system are characterized by the device that they have to assign a power program for. In the distributed case, every agent locally optimizes a local cost function $\cost_\agent$, which is part of the global optimization problem~\eqref{eq:lprh:optim}. Also, local constraints $\constraint_\agent$ have to be taken into account which are represented by constraint~\eqref{eq:lprh:constraint:agent}. The behavior of the agents can be defined by three functions for the costs, the local constraint and the power program; an overview of this is shown in \tabref{lprh:agentbehaviour}. In our model we consider the following types of agents:

\begin{table*}[t!]
\begin{center}
\caption{A summary showing the three functions specifying the behaviors of the agents.\label{tab:lprh:agentbehaviour}}
\begin{tabular}{l l l l}
\toprule
Agent type & Cost & Constraint & Power \\ \midrule

Market Operator &
$\costof{MO} = 0$ &
$\constraintof{MO} = \bar{\power} - \target$ &
$\bar{\power} = \sum_{\agent \in \allagents}{\power_\agent}$ \\

Congestion agent &
$\costof{CO}_\agent = 0$ &
$\constraintof{CO} = \bar{\power_\agent} - \congestionThreshold_\agent \text{ iff } |\bar{\power_\agent}| < \congestionThreshold_\agent$ &
$\bar{\power_\agent} = \sum_{j \in \children_\agent}{\power_j}$ \\

Load agent &
$\costof{LOAD}_\agent = 0$ &
$\constraintof{LOAD}_\agent = 0$ &
$\power_\agent$ \\

PV agent &
$\costof{PV}_\agent = \power_\agent - \power'_\agent$ &
$\constraintof{PV}_\agent = 0 \text{ iff } \slotsize \power_\agent \price \geq \PVMarginalCost_\agent$ &
$\power_\agent = \power'_\agent \text{ or } 0$ \\

Storage agent &
$\costof{ST}_\agent = \power_\agent(1-\storageEfficiency'_i)$ &
$\constraintof{ST}_\agent = 0 \text{ iff } \minenergy_\agent < \energy_\agent < \maxenergy_\agent$ &
$\power_\agent = f(\price)$ \\
\bottomrule
\end{tabular}
\end{center}
\end{table*}

\subsubsection{Market Operator}
This is the root node of the tree ($M$ in \figref{lprh:topology}), as far as the local distribution grid concerns. In the physical grid, it corresponds to the transformer that connects the local low voltage (LV) grid to the medium voltage (MV) grid. We assume that there is no energy loss at the market operator, so
\begin{equation}
\costof{MO} = 0.
\end{equation}
Its goal is to find a solution to~\eqref{eq:lprh:optim}, and its private cost would be the same as the target constraint in~\eqref{eq:lprh:constraint:target}. The market operator runs \algoref{lprh:lprh}, using a deviation of the target profile $\target$ as an error, which is minimized by the algorithm. Hence, its local constraint is defined as
\begin{equation}
\label{eq:lprh:marketoperator_constraint}
\constraintof{MO} = \bar{\power} - \target,
\end{equation}
where $\bar{\power}$ denotes the power program of the market operator. Since the market operator is not a device in the grid itself, its power is the sum of the powers of its children, which in case of the market operator are all nodes in the cluster:
\begin{equation}
\bar{\power} = \sum_{\agent \in \allagents}{\power_\agent}.
\end{equation}

The value of~\eqref{eq:lprh:marketoperator_constraint} can either be negative or positive, which respectively means either the total power production or the total consumption is too high. The market operator sets an initial price of $\price = 0.5$ for all timeslots, and then uses \algoref{lprh:lprh} to minimize the constraint value until it reaches zero, in order to satisfy the global constraint~\eqref{eq:lprh:constraint:target}.

\subsubsection{Congestion Agent}
This is an intermediate node on the grid tree ($C$ in \figref{lprh:topology}) connected to a parent node who is either the market operator or another congestion agent. It corresponds to a component in the grid where congestion might potentially occur, such as a transformer. Equivalently to the market operator, we assume that no energy is lost here, so:
\begin{equation}
\costof{CO}_\agent = 0.
\end{equation}

This agent has a constraint that aims to limit the power usage of that part of the grid, this corresponds to the constraint~\eqref{eq:lprh:constraint:grid}. The congestion agent also uses \algoref{lprh:lprh} to minimize the error of its local constraint, which is defined as
\begin{equation}
\label{eq:lprh:concentrator_cost}
\constraintof{CO}_\agent =
\begin{cases}
\bar{\power_\agent} - \congestionThreshold_\agent, & \text{if } |\bar{\power_\agent}| \geq \congestionThreshold_\agent, \\
0, & \text{otherwise}.
\end{cases}
\end{equation}
Here $\congestionThreshold_{\agent}$ is the power limit of the agent $\agent$, which means it is the maximum power throughput of the grid at the point $\agent$ represents. Again $\bar{\power_{\agent}}$ denotes the power profile of the congestion agent, but is now equal to the sum of all devices under the current node:
\begin{equation}
\bar{\power_\agent} = \sum_{j \in \children_\agent}{\power_j},
\end{equation}
Similar to the constraint of the market operator in~\eqref{eq:lprh:marketoperator_constraint} the constraint value can become negative or positive, and is used to compute the error $\error$ in \algoref{lprh:lprh}.

\subsubsection{Load Agent}
This is an agent responsible for an uncontrollable load in the grid (represented by any device agent $D$ in \figref{lprh:topology}). This could be a consumer household, an office, street lighting, or any other non-flexible load, and thus only participates in the problem as part of constraint~\eqref{eq:lprh:constraint:target}. In the distributed system however, it is responsible for making a forecast of the power usage, and this forecast will be updated with more accurate information as the time horizon shifts. The load agent has no attached cost in the global problem, and no need to locally compute any optimal behavior. This is equivalent to stating that its cost and constraint correspond to
\begin{equation}
\costof{LOAD}_\agent = 0,
\end{equation}
\begin{equation}
\constraintof{LOAD}_{\agent} = 0.
\end{equation}
Its corresponding load profile $\power_{\agent}$ is fixed to some profile that constrains the global problem~\eqref{eq:lprh:optim}. In our experiments its values are taken from real households as described in \secref{lprh:hhw}.

\subsubsection{PV Agent}
The PV agent (any $D$ in \figref{lprh:topology}) has some flexibility to offer to the optimization function by allowing curtailment in reference to the expected generation. We assume that curtailment is binary, in that either the PV generates power as normal, or it is switched off and produces no power at all. Curtailing means that there is potential energy lost, and hence the cost function of a PV agent is defined as
\begin{equation}
\costof{PV}_\agent = \power_\agent - \power'_\agent,
\end{equation}
where $\power'_\agent$ indicates the \emph{expected} power, when not curtailing. This expected power is taken from the scenario, which will be detailed in the \secref{lprh:experiments}.

When reacting to prices in the distributed system, a decision is made in order to decide whether to curtail based on the price profile. If the operational running costs $\PVMarginalCost_{\agent}$ of the PV is \emph{more} than the power that would be generated by it, there is no point in running the generator (from an economic point of view). Hence, the local constraint and its corresponding decision rule of a PV agent can be written as
\begin{equation}
\constraintof{PV}_{\agent} =
\begin{cases}
1 & \text{if } \slotsize \power_\agent \price < \PVMarginalCost_\agent, \\
0 & \text{otherwise},
\end{cases}
\end{equation}
\begin{equation}
\power_\agent =
\begin{cases}
0 & \text{if } \slotsize \power'_\agent \price < \PVMarginalCost_\agent, \\
\power'_\agent & \text{otherwise}.
\end{cases}
\end{equation}
For the PV agent it also holds that the corresponding expected power profile $\power'_{\agent}$ determines the global problem~\eqref{eq:lprh:optim}. Its values in our experiments are taken from real PV panels as described in \secref{lprh:hhw}.

\subsubsection{Storage Agent}
A storage agent (again a leaf node $D$ in \figref{lprh:topology}) provides flexibility by allowing to store some energy in a local storage like a battery or a heat buffer. There are limits to the amount of energy that can be stored, either because of the physical limitations of the storage device, or because of the end-user settings. Moreover, a storage agent has some efficiency, which defines energy loss when energy is put into it, or out from it. This means that we can define the cost function of the storage agent as the energy lost during charging or discharging
\begin{equation}
\costof{ST}_\agent = \power_\agent(1-\storageEfficiency'_i),
\end{equation}
where
\begin{equation}
\storageEfficiency'_\agent =
\begin{cases}
\storageEfficiency_\agent & \text{if } \power_\agent \geq 0, \\
\storageEfficiency_{\agent}^{-1} & \text{otherwise}.
\end{cases}
\end{equation}
This difference makes sure that the loss is correlated to the internal power of the battery when charging or discharging. I.e. $\power_\agent$ defines the power at the grid side of the storage, when charging a lower power effectively charges the battery, and conversely when discharging a higher power is required to provide some power level to the grid.

In order to define the constraints of the storage agent we must define the update function for the amount of energy $\energy_{\agent}$ stored as the cumulative sum of the powers
\begin{equation}
e_{\agent{}\timeslot} = e_{\agent0} + \slotsize \sum_{\tindex=\timeslot_0}^{\tindex = \timeslot}{\storageEfficiency'_\agent \power_{\agent\tindex} - \leakage_\agent}.
\end{equation}
Here $\leakage_\agent$ represents the leakage or self-discharge rate of the storage agent $\agent$, and $e_{\agent0}$ is the stored energy at the start of the experiment. Let us now define the following constraints on the storage agent
\begin{equation}
\constraintof{ST}_{\agent} =
\begin{cases}
0, & \text{if } \minenergy_\agent < \energy_{\agent} < \maxenergy_\agent, \\
1, & \text{otherwise.}
\end{cases}
\end{equation}
The minimum and maximum energy values are defined by $\minenergy_\agent$ and $\maxenergy_\agent$, respectively.

For a storage device, the power limits denote the maximum charge and discharge rates. They are defined by $\maxpower_\agent$ and $\minpower_\agent$, respectively, in~\eqref{eq:lprh:constraint:power}. A special case of a storage device is a heat pump, which (electrically powered) heats a house in order to keep the temperature within comfortable levels. The heat pump only allows charging and only discharges through leakage, hence for a heat pump $\minpower_\agent = 0$ and $\leakage_\agent > 0$.

When a storage agent has to update its expected power profile in line~\ref{line:lprh:localoptimization} of \algoref{lprh:lprh}, some response function ($f(\price)$ in \tabref{lprh:agentbehaviour}) is required, which is ``economically sane'' and has the following characteristics:

\begin{itemize}
\item return $\maxpower_\agent$ for low prices and $\minpower_\agent$ for high prices,
\item be a monotonically decreasing for increasing prices,
\item have a ``plateau'' of zero power response for some intermediate price ($\price = 0.5$), which is wider for less efficient devices.
\end{itemize}

\begin{figure}[t]
\centering
\begin{tikzpicture}

\begin{axis}[
height=0.25\figurewidth,
xmin=0,
xmax=1,
xlabel={Price (steering signal)},
ymin=-120,
ymax=220,
ylabel={Power (W)}
]
\addplot [color=plotcolor1] table[]{
0.0 200
0.375 200
0.45 0
0.55 0
0.625 -100
1.0 -100
};

\end{axis}
\end{tikzpicture}
\caption{The strategy of the storage agent with $\maxpower_\agent = 200$\,W, $\minpower_\agent = -100$\,W and $\storageEfficiency_\agent = 0.9$ shows the power response for an increasing price. The \emph{plateau} at $0$\,W extends from $\price = \storageEfficiency_\agent / 2$ to $\price = 0.5 / \storageEfficiency_\agent$.}
\label{fig:lprh:storage_bidcurve}
\end{figure}
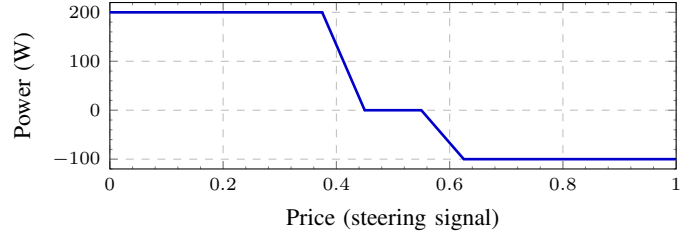

The final characteristic allows more efficient devices to respond to subtle price change, and have less efficient devices respond to more extreme prices. This way devices with a higher efficiency are used first when flexibility is needed, and (when parameterized correctly) less efficient devices will only be used when required. In our implementation we chose the simplest response, where the agent linearly decreases its power from $\maxpower_\agent$ to zero at $\price = \storageEfficiency_\agent / 2$. It then stays at zero symmetrically around $\price = 0.5$ until $\price = 0.5 / \storageEfficiency_\agent$, and then decreases its power response linearly to $\minpower_\agent$. This power/price relation is depicted in \figref{lprh:storage_bidcurve}.

An alternative strategy for the storage agent would be to make use out of any differences in the price, charging and discharging as soon as the price differences are large enough to overcome its efficiency. From a strictly economic perspective, this is the optimal strategy to maximize its own benefit. However, this leads to very ``binary'' behavior with minimal and maximal charging rates~\cite{Li2011} and thus, little room for optimizing from the market operator and congestion agent. Therefore, a linear strategy is implemented as depicted in \figref{lprh:storage_bidcurve}, allowing to solve the overall optimization problem~\eqref{eq:lprh:optim}.

\section{Experiment Setup}
\label{sec:lprh:experiments}
The LP-RH algorithm was empirically evaluated by running a simulation of an LV grid with a set of realistic household load profiles and PV production profiles for a series of 24 PTUs.

\subsection{Distribution Network Topology}
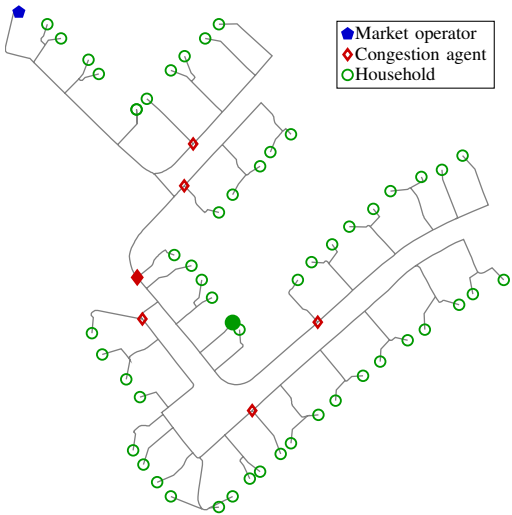
\begin{figure}[t]
\centering
\begin{tikzpicture}
\begin{axis}[
width=.75\figurewidth,
height=.75\figurewidth,
unbounded coords=jump,
axis lines=none,
xmin=-.1,
xmax=145.5,
ymin=-.1,
ymax=140.5,
ticks=none,
minor tick num=0,
xmajorgrids=false,
ymajorgrids=false,
legend style={legend cell align=left, align=left}
]
\addplot [color=gray, line width=.5pt, forget plot] table[]{
3.9 140.1
2.8 140.3
2.7 140.3
2.6 140.3
2.5 140.3
2.4 140.1
0 130.3
0 129.9
0 129.5
0 129.2
0.1 129.0
0.3 128.8
0.4 128.7
0.7 128.6
3.5 127.9
6.9 127.1
7.1 127.1
7.3 126.9
7.6 126.6
8.2 126.0
11.6 130.9
12.5 132.2
16.2 132.7
nan nan
8.2 126.0
15.4 119.1
17.3 117.3
21.4 121.3
21.6 121.1
21.8 120.9
22.0 120.9
22.1 120.9
22.2 120.9
22.4 120.9
23.2 121.7
27.2 122.7
nan nan
11.6 130.9
10.4 132.0
10.4 132.1
10.4 132.1
10.4 132.2
10.4 132.5
10.4 132.7
11.1 133.4
12.2 136.7
nan nan
17.3 117.3
23.1 111.8
27.2 107.9
30.2 105.1
30.7 104.6
32.7 102.8
33.1 102.4
38.1 98.1
39.8 96.6
40.2 96.3
41.1 95.7
41.9 95.3
42.4 95.0
42.8 94.9
43.2 94.6
43.5 94.6
43.8 94.7
44.2 94.7
44.8 94.8
45.6 95.0
46.1 95.3
46.8 95.6
47.6 96.1
48.3 96.5
49.3 97.4
53.7 102.0
54.7 103.0
46.3 111.2
44.7 112.7
41.2 115.7
nan nan
21.4 121.3
22.6 122.7
24.2 126.7
nan nan
32.7 102.8
38.4 108.7
38.2 112.7
38.4 108.7
nan nan
43.2 94.6
43.5 94.4
43.6 94.3
48.7 88.7
49.2 88.2
50.0 89.0
50.3 89.4
52.1 91.2
54.8 88.5
55.0 88.3
55.2 88.0
55.3 87.8
55.3 87.4
55.4 86.9
55.4 86.6
55.6 86.2
55.8 85.9
56.1 85.6
56.6 85.1
57.8 83.9
58.0 83.8
58.0 83.8
58.1 83.7
58.1 83.7
58.2 83.8
58.4 83.9
58.5 84.0
58.9 84.5
62.2 83.7
nan nan
49.2 88.2
46.7 85.5
42.1 80.8
40.0 78.7
39.5 78.0
38.8 77.3
38.3 76.5
37.7 75.6
37.6 75.3
36.7 73.7
36.4 72.8
36.2 72.0
36.1 71.4
36.2 70.5
36.5 69.0
36.9 67.7
37.3 66.9
37.6 66.4
38.4 65.4
42.0 68.2
42.5 68.6
42.7 68.8
42.9 69.1
43.1 69.6
43.2 70.0
43.4 70.4
43.7 70.7
44.0 71.1
44.4 71.4
44.6 71.7
45.7 72.6
46.2 72.9
46.3 73.0
46.5 73.0
46.6 73.1
46.7 73.0
46.7 73.0
46.8 72.9
47.1 72.4
49.2 71.7
nan nan
52.1 91.2
57.0 96.4
60.7 100.3
63.2 102.8
65.7 105.4
72.2 112.2
74.5 114.6
77.5 111.6
77.8 111.3
78.0 110.9
78.2 110.6
78.2 110.3
78.3 109.7
78.4 109.3
78.5 109.0
78.7 108.7
79.0 108.4
80.3 107.1
80.6 106.8
80.7 106.7
80.8 106.7
80.3 106.0
80.0 105.7
79.6 105.2
78.7 104.3
77.2 100.7
nan nan
63.2 102.8
66.8 99.1
67.0 98.9
67.2 98.7
67.3 98.6
67.3 98.4
67.4 97.7
67.5 97.3
67.6 97.1
67.8 96.6
67.9 96.4
68.1 96.3
69.6 94.8
70.0 94.4
70.1 94.3
67.9 91.7
66.2 88.7
nan nan
54.7 103.0
58.4 106.8
63.2 112.0
64.0 112.8
67.1 116.2
72.8 122.3
70.3 124.5
62.7 131.9
58.2 132.7
nan nan
63.2 112.0
61.4 113.6
54.2 120.6
52.8 122.0
48.2 123.7
nan nan
54.2 120.6
55.0 121.4
55.1 121.6
55.2 121.7
55.2 121.8
55.1 121.9
55.0 122.1
54.8 122.3
54.1 123.0
52.2 127.7
nan nan
72.8 122.3
74.2 123.9
76.8 126.8
77.7 127.8
68.8 135.6
67.5 136.8
62.2 136.7
nan nan
38.4 65.4
39.0 64.7
41.1 62.3
44.4 65.0
44.6 65.2
44.8 65.3
45.2 65.4
46.8 65.7
47.1 65.8
47.3 65.9
47.5 66.1
47.7 66.2
49.6 67.7
54.2 68.7
nan nan
70.1 94.3
70.2 94.3
70.4 94.3
70.5 94.3
70.6 94.4
70.7 94.4
71.3 95.0
74.2 96.7
nan nan
41.1 62.3
41.8 61.5
44.4 58.6
45.5 57.2
46.0 56.6
49.5 52.6
49.7 52.3
50.3 52.7
52.8 54.8
53.8 55.5
55.7 57.1
57.4 58.4
60.2 59.7
nan nan
44.4 58.6
43.7 58.0
40.5 55.8
39.9 55.2
39.7 55.0
39.7 54.9
39.6 54.8
39.6 54.6
39.5 54.5
39.5 54.4
39.5 54.4
39.5 54.3
39.5 54.2
39.6 54.1
39.9 53.7
42.2 50.9
44.4 48.1
44.5 48.0
41.3 45.6
40.9 45.4
40.8 45.3
40.6 45.2
40.4 45.1
39.7 44.9
38.8 44.6
38.4 44.5
38.0 44.3
37.6 44.1
37.0 43.7
36.6 43.4
36.3 43.2
36.2 43.0
36.2 42.9
36.1 42.9
36.1 42.8
36.1 42.7
36.3 42.2
35.1 43.7
33.7 45.3
28.2 43.7
nan nan
80.8 106.7
80.9 106.7
81.0 106.7
81.1 106.8
81.2 106.9
81.7 107.3
83.2 105.7
nan nan
49.7 52.3
54.5 46.9
56.7 44.2
63.8 50.0
66.2 52.7
nan nan
56.7 44.2
60.9 39.1
66.2 43.4
68.8 45.6
69.1 45.8
69.1 45.9
69.2 46.0
69.2 46.1
69.2 46.2
69.2 46.3
69.2 46.4
69.1 46.5
69.0 46.6
68.5 47.0
68.2 50.7
nan nan
60.9 39.1
61.7 38.1
62.4 37.3
62.9 36.8
63.7 36.3
64.2 36.0
64.9 35.6
65.8 35.4
66.8 35.2
67.5 35.2
68.4 35.3
69.6 35.6
70.4 35.9
71.2 36.3
71.9 36.8
72.7 37.5
74.1 38.6
78.6 42.4
85.0 47.8
89.7 51.6
91.0 52.8
95.1 56.4
99.6 60.3
101.2 61.8
104.7 64.8
109.8 69.1
113.1 71.5
113.6 72.0
115.1 73.2
116.2 74.0
117.8 75.0
118.9 75.7
122.2 77.5
124.6 78.8
125.1 79.0
125.3 79.1
128.7 80.8
131.2 81.9
136.2 84.0
140.7 85.8
139.2 89.4
137.3 94.5
136.8 95.9
133.2 99.7
nan nan
55.7 57.1
55.1 57.9
55.0 58.0
55.0 58.1
54.9 58.3
55.0 58.4
55.1 58.6
55.2 58.7
56.4 59.7
57.2 64.7
nan nan
39.9 53.7
38.5 53.9
30.0 55.5
26.9 56.1
26.5 56.1
26.3 56.1
26.1 56.1
25.3 55.6
25.1 55.4
25.0 55.2
25.0 55.0
25.0 54.7
25.6 54.1
25.2 49.7
nan nan
44.5 48.0
46.3 45.8
47.2 44.7
49.8 41.4
53.6 36.8
54.4 35.8
54.6 35.6
54.7 35.4
54.8 35.3
54.8 35.2
54.8 35.0
54.8 34.9
54.7 34.8
54.7 34.6
54.6 34.5
54.3 34.2
47.4 28.8
43.9 33.1
39.2 31.7
nan nan
91.0 52.8
88.5 55.5
88.1 55.9
87.9 56.1
87.7 56.2
87.4 56.3
87.0 56.4
86.7 56.5
86.5 56.6
86.4 56.8
86.0 57.2
82.9 60.5
82.7 60.7
82.7 60.8
82.6 61.0
82.6 61.2
82.7 61.4
82.8 61.6
82.9 61.8
83.0 61.9
83.6 62.3
85.2 64.7
nan nan
101.2 61.8
94.2 69.3
93.9 69.8
93.7 69.9
93.6 70.0
93.5 70.0
93.3 70.0
93.2 70.0
92.9 69.7
92.7 69.5
89.2 69.7
nan nan
94.2 69.3
96.7 72.6
95.2 74.7
nan nan
47.4 28.8
46.4 28.0
45.8 27.6
45.7 27.5
45.7 27.4
45.7 27.3
45.6 27.3
45.7 27.2
45.7 27.1
45.9 26.8
42.2 23.9
42.0 23.7
41.9 23.6
41.7 23.6
41.7 23.6
41.5 23.6
41.2 23.6
40.9 23.7
40.6 23.7
40.4 23.7
40.3 23.7
40.0 23.7
39.7 23.7
39.6 23.7
39.4 23.6
39.3 23.6
39.1 23.5
38.9 23.4
38.3 22.9
37.6 22.3
37.5 22.2
37.5 22.1
37.5 22.1
37.4 22.0
37.4 21.9
37.5 21.9
37.5 21.8
37.6 21.7
37.9 21.3
37.2 16.7
nan nan
113.6 72.0
110.4 75.4
108.4 77.4
106.9 79.1
106.3 79.7
106.1 79.9
106.0 80.0
105.9 80.0
105.8 80.0
105.7 80.0
105.6 80.0
105.1 79.6
100.2 79.7
nan nan
36.3 42.2
35.2 36.7
nan nan
106.3 79.7
109.0 82.5
107.2 84.7
nan nan
124.6 78.8
122.3 81.8
122.0 82.2
121.9 82.6
121.7 82.9
120.8 86.0
120.3 87.5
120.0 88.1
119.8 88.5
119.3 89.1
118.4 90.2
118.1 90.5
118.1 90.6
118.0 90.6
117.9 90.6
117.8 90.6
117.2 90.2
112.2 89.7
nan nan
45.9 26.8
46.4 26.2
48.8 23.1
41.7 17.3
40.2 12.7
nan nan
131.2 81.9
131.8 82.2
127.4 93.0
127.2 95.7
nan nan
48.8 23.1
51.7 19.4
54.6 16.0
53.4 15.0
53.1 14.7
52.9 14.5
52.7 14.2
52.3 13.4
52.0 12.7
51.7 12.3
51.4 11.9
51.1 11.6
50.6 11.2
49.0 9.7
44.2 7.7
nan nan
54.6 16.0
55.4 15.1
55.6 14.9
55.7 14.8
55.8 14.7
56.0 14.6
56.2 14.6
56.4 14.6
56.6 14.7
56.9 14.3
57.1 14.1
57.3 13.9
57.4 13.7
57.4 13.5
57.4 13.2
57.4 12.6
56.8 8.7
56.3 5.6
56.2 4.6
56.1 3.6
56.1 3.1
56.1 2.7
56.2 2.4
56.3 2.0
56.4 1.7
56.7 1.1
57.3 0.2
57.6 0.1
57.9 0
58.2 0
58.4 0
58.6 0
58.7 0
58.8 0.1
59.0 0.2
59.1 0.3
59.4 0.5
62.2 0.7
nan nan
118.1 90.6
118.5 90.9
119.1 91.2
121.7 92.6
121.2 93.7
nan nan
56.6 14.7
57.3 15.2
61.2 18.6
63.9 21.0
69.3 14.9
68.7 14.4
68.6 14.2
68.5 14.1
68.4 14.0
68.4 14.0
68.4 13.9
68.4 13.8
68.4 13.7
68.4 13.6
68.5 13.5
68.6 13.4
69.4 12.4
71.2 8.7
nan nan
63.9 21.0
71.9 27.9
78.2 20.8
80.2 15.7
nan nan
69.3 14.9
70.7 13.3
74.2 10.7
nan nan
71.9 27.9
73.6 29.3
75.3 30.9
80.1 35.1
85.8 40.0
91.7 45.0
94.2 47.1
97.6 49.9
101.5 53.3
104.7 55.9
105.9 56.9
107.5 58.4
111.5 61.7
114.5 64.7
117.0 66.9
119.4 68.6
119.6 68.6
119.9 68.7
120.1 68.6
120.2 68.6
120.4 68.4
120.6 68.3
120.7 68.2
120.8 68.1
121.8 66.1
122.6 64.8
123.2 63.7
123.8 62.5
125.0 60.5
125.3 60.0
125.6 59.5
126.2 58.7
126.3 58.5
126.5 58.4
126.6 58.4
126.8 58.5
127.1 58.6
127.6 59.0
nan nan
80.1 35.1
82.7 32.1
83.0 31.7
83.1 31.6
83.2 31.5
83.2 31.3
83.3 31.2
83.3 31.0
83.3 30.8
83.4 30.5
83.4 30.2
83.4 29.9
83.5 29.7
83.5 29.6
83.6 29.5
83.6 29.3
83.8 29.1
85.3 27.3
85.9 26.6
86.3 26.2
86.3 26.2
86.4 26.1
86.5 26.1
86.6 26.1
86.6 26.1
86.7 26.1
86.8 26.1
88.0 26.4
83.6 22.6
83.2 18.7
nan nan
91.7 45.0
97.8 37.9
98.6 36.9
99.0 36.4
96.4 33.7
96.2 30.7
nan nan
99.0 36.4
99.1 36.3
99.1 36.3
99.2 36.2
99.3 36.2
99.5 36.3
99.6 36.3
99.8 36.5
100.9 37.5
104.2 37.7
nan nan
104.7 55.9
108.6 51.5
111.9 47.8
109.4 45.2
109.2 41.7
nan nan
111.9 47.8
112.0 47.7
112.2 47.7
112.3 47.7
112.4 47.7
112.5 47.7
112.6 47.7
112.7 47.8
113.6 48.5
117.2 47.7
nan nan
57.3 0.2
55.4 0.9
52.3 2.1
49.7 3.1
48.2 3.7
nan nan
119.4 68.6
121.8 70.1
124.5 71.7
126.8 72.8
129.2 74.2
131.2 75.2
133.5 76.1
135.5 71.7
135.8 71.1
136.5 69.2
137.5 67.1
137.7 66.9
138.0 66.7
138.3 66.6
138.8 66.6
139.5 66.8
141.8 68.1
145.2 64.7
nan nan
59.4 0.5
66.2 3.7
nan nan
137.7 66.9
137.0 66.4
135.3 65.3
136.2 60.7
nan nan
125.3 60.0
123.5 57.4
122.0 55.2
122.2 50.7
nan nan
88.0 26.4
91.2 26.7
nan nan
127.6 59.0
132.2 57.7
};

\addplot[only marks, mark size=2pt, color=plotcolor1, mark=pentagon*] table[]{
3.9 140.1
};
\addlegendentry{Market operator}

\addplot[only marks, mark size=2.5pt, color=plotcolor2, mark=diamond*, forget plot] table[]{
38.40 65.40
};
\addplot[only marks, mark size=2pt, color=plotcolor2, mark=diamond] table[]{
52.10 91.2
54.7 103
38.40 65.40
39.90 53.7
91 52.8
71.90 27.90
};
\addlegendentry{Congestion agent}

\addplot[only marks, mark size=2.5pt, color=plotcolor3, mark=*, forget plot] table[]{
66.2 52.7
};
\addplot[only marks, mark size=2pt, color=plotcolor3, mark=o, line width=.7pt] table[]{
16.2 132.7
24.2 126.7
12.2 136.7
38.2 112.7
38.2 112.7
27.2 122.7
41.2 115.7
48.2 123.7
58.2 132.7
62.2 136.7
52.2 127.7
66.2 88.7
62.2 83.7
74.2 96.7
77.2 100.7
83.2 105.7
66.2 52.7
54.2 68.7
60.2 59.7
49.2 71.7
57.2 64.7
68.2 50.7
25.2 49.7
95.2 74.7
39.2 31.7
35.2 36.7
28.2 43.7
107.2 84.7
89.2 69.7
100.2 79.7
40.2 12.7
127.2 95.7
85.2 64.7
133.2 99.7
121.2 93.7
112.2 89.7
44.2 7.74
74.2 10.7
80.2 15.7
96.2 30.7
71.2 8.74
109.2 41.7
37.2 16.7
104.2 37.7
48.2 3.74
117.2 47.7
62.2 0.742
66.2 3.74
136.2 60.7
91.2 26.7
122.2 50.7
145.2 64.7
83.2 18.7
132.2 57.7
};
\addlegendentry{Household}

\end{axis}
\end{tikzpicture}
\caption{The topology of the IEEE LV feeder network used for the simulation. The figure depicts the connections between household consumers, the intermediate congestion agents, and the root market operator. The ``filled'' markers represent the agents from \secref{lprh:results}.}
\label{fig:lprh:topology}
\end{figure}

For the topology of the network we use the European Low Voltage Test Feeder~\cite{Feeder2015} network. This dataset is used to benchmark power and energy algorithms on realistic European distribution networks. In this paper{} we superimposed six points on the topology, where we monitor and mitigate any potential congestion. These points are strategically chosen to separate the problem into independent subproblems. The resulting topology with the congestion points are shown in \figref{lprh:topology}.

\subsection{Household Load and PV Profiles}
\label{sec:lprh:hhw}
The household consumption and production profiles are taken from a pilot study~\cite{Rassa2019}, in which 92 residential consumers were monitored over the course of a year (from March through November 2018). The data was preprocessed such that we have separated information on the consumption of houses, and of the PV installations. Data is anonymized and randomized per month, so that we can select data from any specific month for a base load of a household, or a residential PV installation.

In the experiment the 54 households from \figref{lprh:topology} were assigned a random instance of the load \emph{and} PV profiles from the same month (i.e. all households were equipped with PV panels). The daily load consumption varied between 4.98\,kWh and 29.39\,kWh, and the total PV production varied between 826\,Wh and 18.8\,kWh. Furthermore, 16 randomly selected households were assigned a household battery, and again 16 were chosen to have a heat pump installation.

Every household and PV installation in the simulation would select a random profile from the dataset, which was used as its power profile. The objective $\target$ was set to a total net consumption of the LV grid using the mean total power consumption of the houses including PV production. Choosing the target profile $\target$ in this way corresponds to a situation in which the energy provider of the simulated neighborhood would agree to a contract for the average behavior of the households, and subsequently attempts to use flexibility to account for any deviations from the normal.

The batteries were all dimensioned with storage capacities of $\maxenergy = 10.8\text{\,kWh}$, and maximum charge and discharge rates of $\maxpower = 4000\text{\,W}$ and $\minpower = -4000\text{\,W}$, respectively. The batteries charging efficiencies were all set at $\storageEfficiency=0.9$. The heat pumps were estimated to have a working energy capacity of $\maxenergy = 2\text{\,kWh}$, this means the difference between the thermal energy capacity of the house at the minimum and maximum comfortable user temperature is $2 \cop\text{\,kWh}$, where $\cop = 4.0$ is the coefficient of performance of the heat pump. Then, the heat pumps have a $\maxpower = 1600\text{\,W}$ and $\minpower = 0\text{\,W}$, an efficiency of $\storageEfficiency = 1$ and a constant leakage rate of $\leakage = 360\text{\,W}$. Finally, to ensure convergence, the maximal error value is set to $\errorbound=10^{-3}$ in the experiments for this paper{}

\subsection{Forecast Uncertainty}
The predicted load and production power profiles $\hat{\power}$ of the load and PV agents were generated by taking average profiles of the complete dataset. These average profiles are considered as taken from historic data and hence, provide a ground for predicting the power program of future PTUs. When agent $\agent$ determines its power prediction $\power'_\agent$, it will compute a weighted average between its assigned power profile $\power_\agent$ and the average profile $\hat{\power}$, such that:
\begin{align}
\label{eq:lprh:forecast}
\power_{\agent{}\timeslot} &= (1-\alpha) \power'_{\agent{}\timeslot} + \alpha \hat{\power}_{\timeslot}, \\
\alpha &= \sqrt{\frac{\timeslot-1}{\horizon-1}},
\end{align}
such that at $\timeslot = 1$ the prediction equals the selected profile $\power'_{\agent{}\timeslot}$. At $\timeslot = \horizon$, the prediction is simply the average power $\hat{\power}_\timeslot$.

\section{Centralized Solver}
In order to address the performance of the LP-RH algorithm, a centralized optimization approach is also introduced to provide lower bounds to its results. A mixed integer linear program (MILP) solver was used to find these bounds for the problem stated in~\eqref{eq:lprh:optim}. The centralized optimization approach considers the exact same scenario that was solved by the decentralized algorithm; the energy loss is minimized and the same set of constraints apply. A fundamental difference lies in the availability of information. Whereas detailed information is only shared locally in the LP-RH algorithm, the centralized optimization approach assumes complete knowledge of the current system state for the decision-maker; i.e., no limits are imposed on the spacial flow of information within the network. With these features in mind, two versions of the MILP were formulated: Receding Horizon Centralized Solver and Perfect Information Centralized Solver.

\subsection{Receding Horizon Centralized Solver}
The receding horizon centralized solver (RHCS) is the most similar to the LP-RH algorithm. It uses a receding horizon approach (as depicted in \figref{lprh:receding-horizon}) to find an ideal dispatch solving the consecutive sub-problems. Uncertainty about future device states is again simulated by~\eqref{eq:lprh:forecast}. Because this solver has perfect information at the \emph{current} state for each iteration of the receding horizon, its solution represents a lower bound on the solution found by the LP-RH algorithm.

\subsection{Perfect Information Centralized Solver}
In the perfect information centralized solver (PICS), a single centralized optimization problem is solved for the complete dispatch. In addition to perfect spacial information, this version of the MILP also has perfect \emph{temporal} information about the complete system state; this means that no uncertainty about future states is simulated. Referring back to~\eqref{eq:lprh:forecast}, this is equivalent to setting $\alpha = 1$, which results in perfect predictions for each profile. The solution of this MILP represents an absolute lower bound for problem~\eqref{eq:lprh:optim}.

\section{Results}
\label{sec:lprh:results}
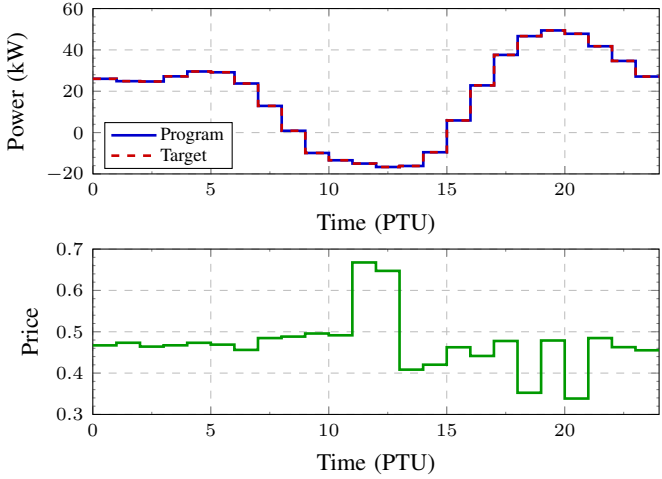
\begin{figure}[!t]
\centering
\begin{tikzpicture}
\begin{axis}[
height=\subfigheight,
at={(0,\subfigtop)},
scale only axis,
minor x tick num=1,
xmin=0,
xmax=24,
xlabel={Time (PTU)},
ymin=-20,
ymax=60,
ylabel={Power (kW)},
legend style={at={(0.02,0.03)}, anchor=south west, legend cell align=left, align=left, draw=black}
]
\addplot[const plot, color=plotcolor1] table[] {
0 26.023712579752
1 24.869503942937
2 24.7142007991772
3 27.2083582704921
4 29.5652362151693
5 29.1686574750236
6 23.7451653542953
7 12.9080432862165
8 0.87379722641341
9 -9.86505710319685
10 -13.3949232159685
11 -14.9884842414685
12 -16.6507204607126
13 -16.154114163685
14 -9.50886612487795
15 5.8759133642362
16 22.8290905105984
17 37.5743503231772
18 46.6590471786614
19 49.4241613553032
20 47.8074802929331
21 41.7641929311968
22 34.6604527980118
23 27.1122520047165
24 27.1122520047165
};
\addlegendentry{Program}

\addplot[const plot, color=plotcolor2, dashed] table[] {
0 26.023712579752
1 24.869503942937
2 24.7142007991772
3 27.2083582704921
4 29.5652362151693
5 29.1686574750236
6 23.7451653542953
7 12.9080432862165
8 0.873797226413384
9 -9.86505710319685
10 -13.3949232159685
11 -14.9884842414685
12 -16.6507204607126
13 -16.154114163685
14 -9.50886612487795
15 5.87591336423621
16 22.8290905105984
17 37.5743503231772
18 46.6590471786614
19 49.4241613553032
20 47.8074802929331
21 41.7641929311969
22 34.6604527980118
23 27.1122520047165
24 27.1122520047165
};
\addlegendentry{Target}

\end{axis}

\begin{axis}[
height=\subfigheight,
at={(0,0)},
scale only axis,
minor x tick num=1,
xmin=0,
xmax=24,
xlabel={Time (PTU)},
ymin=0.3,
ymax=0.7,
ylabel={Price}
]
\addplot[const plot, color=plotcolor3] table[] {
0 0.46709694378946
1 0.473389001706872
2 0.464086544732755
3 0.467215921162545
4 0.47330391551742
5 0.468862484498076
6 0.456109660908683
7 0.484714343938425
8 0.488289896655922
9 0.495788878972211
10 0.491443668231034
11 0.667638140553775
12 0.647266715479071
13 0.408387322136327
14 0.420318480949411
15 0.462489293852867
16 0.441454668790988
17 0.477668463166289
18 0.352458819662482
19 0.478846518990526
20 0.338452409886032
21 0.484738346152724
22 0.462731164242489
23 0.455195037925993
24 0.455195037925993
};

\end{axis}
\end{tikzpicture}
\caption{The power program of the market operator and the price profile as the outcome of \algoref{lprh:lprh}, show the results of a 24-hour simulation of the problem with the LV feeder network and 92 random households.}
\label{fig:lprh:result_mo}
\end{figure}

In this section the results of the described experiments are shown. Before comparing the performance of the different solvers, which is done in \secref{lprh:comparison}, the behavior of the LP-RH algorithm is shown in this section. The graphs in this section show examples of single simulation runs, demonstrating the behavior of the different agent types. Powers are shown as power consumption, this means a net consumption is shown as positive power, and conversely negative powers indicates a net power production. In \figref{lprh:result_mo} the final result of the market operator is shown, where the market operator has found a price profile such that the target profile is met exactly.

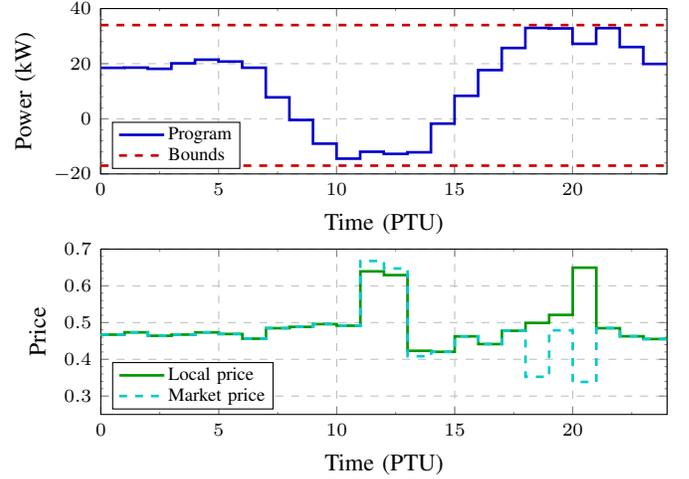
\begin{figure}[!t]
\centering
\begin{tikzpicture}

\begin{axis}[
height=\subfigheight,
at={(0,\subfigtop)},
scale only axis,
minor x tick num=1,
unbounded coords=jump,
xmin=0,
xmax=24,
xlabel={Time (PTU)},
ymin=-20,
ymax=40,
ylabel={Power (kW)},
legend style={at={(0.02,0.03)}, anchor=south west, legend cell align=left, align=left, draw=black}
]
\addplot [const plot, color=plotcolor1] table[] {
0 18.4923110221788
1 18.5732430197053
2 18.1386132190092
3 20.1291364120452
4 21.4547275210856
5 20.7812315249165
6 18.5325613583834
7 7.78485035786654
8 -0.429344066773682
9 -9.01844583204724
10 -14.4594785932163
11 -11.9526279858937
12 -12.7295760503527
13 -12.1562426451617
14 -1.74481725400864
15 8.33472137763603
16 17.6787343448804
17 25.6921577926651
18 32.9689775660621
19 32.8014166648333
20 27.2062086662785
21 32.8940699807507
22 26.041992402458
23 19.9080853306332
24 19.9080853306332
};
\addlegendentry{Program}

\addplot [color=plotcolor2, dashed] table[] {
0 -17
24 -17
nan nan
0 34
24 34
};
\addlegendentry{Bounds}

\end{axis}

\begin{axis}[
height=\subfigheight,
at={(0,0)},
scale only axis,
minor x tick num=1,
xmin=0,
xmax=24,
xlabel={Time (PTU)},
ymin=0.25,
ymax=0.7,
ytick={0.3, 0.4, 0.5, 0.6, 0.7},
ylabel={Price},
legend style={at={(0.02,0.03)}, anchor=south west, legend cell align=left, align=left, draw=black}
]
\addplot[const plot, color=plotcolor3] table[] {
0 0.46709694378946
1 0.473389001706872
2 0.464086544732755
3 0.467215921162545
4 0.47330391551742
5 0.468862484498076
6 0.456109660908683
7 0.484714343938425
8 0.488289896655922
9 0.495788878972211
10 0.491443668231034
11 0.639397694868279
12 0.629218402522711
13 0.423118660178081
14 0.420318480949411
15 0.462489293852867
16 0.441454668790988
17 0.477668463166289
18 0.499066403215792
19 0.520857294471435
20 0.649399989330777
21 0.484738346152724
22 0.462731164242489
23 0.455195037925993
24 0.455195037925993
};
\addlegendentry{Local price}

\addplot[const plot, color=plotcolor4, dashed] table[] {
0 0.46709694378946
1 0.473389001706872
2 0.464086544732755
3 0.467215921162545
4 0.47330391551742
5 0.468862484498076
6 0.456109660908683
7 0.484714343938425
8 0.488289896655922
9 0.495788878972211
10 0.491443668231034
11 0.667638140553775
12 0.647266715479071
13 0.408387322136327
14 0.420318480949411
15 0.462489293852867
16 0.441454668790988
17 0.477668463166289
18 0.352458819662482
19 0.478846518990526
20 0.338452409886032
21 0.484738346152724
22 0.462731164242489
23 0.455195037925993
24 0.455195037925993
};
\addlegendentry{Market price}

\end{axis}
\end{tikzpicture}
\caption{The power program of a congestion agent shows some periods of congestion at the production and consumption bounds (red dashed lines), and the corresponding changes in price profiles (from \algoref{lprh:lprh}) relative to the global price of the market operator.}
\label{fig:lprh:result_ag}
\end{figure}

\figref{lprh:result_ag} shows the power profile of one of the three congestion agents that is directly connected to the market operator. Its power congestion limits are set such that in the peak moments of the day there is some congestion expected. This results in a price difference shown in the power program, as around the peak PV production ($\timeslot = {12, 13, 14}$) the local price is slightly lower than the market price, leading to a lower net power production. Similarly, at the end of the day ($\timeslot = {19, 20, 21, 22}$) a positive power congestion was mitigated by increasing the price.

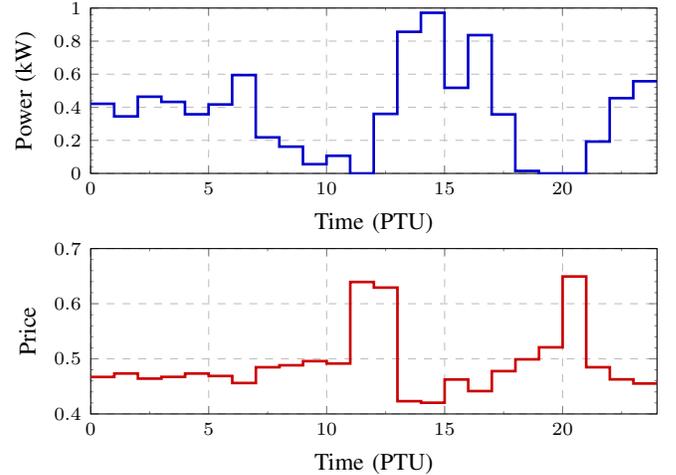
\begin{figure}[!t]
\centering
\begin{tikzpicture}

\begin{axis}[
height=\subfigheight,
at={(0,\subfigtop)},
scale only axis,
minor x tick num=1,
xmin=0,
xmax=24,
xlabel={Time (PTU)},
ymin=0,
ymax=1,
ylabel={Power (kW)}
]
\addplot[const plot, color=plotcolor1] table[] {
0 0.421159119494914
1 0.344838785449188
2 0.463960469333989
3 0.432610935640133
4 0.35762934466683
5 0.41692130030356
6 0.594760331622622
7 0.217971455841658
8 0.161862336218546
9 0.0558193425501972
10 0.106694359290595
11 -0
12 0.36
13 0.856352385816619
14 0.971463015099353
15 0.517616978181351
16 0.836298435113043
17 0.35694395053909
18 0.0149110908488512
19 -0
20 -0
21 0.192568998087333
22 0.455225128461276
23 0.557406480763013
24 0.557406480763013
};

\end{axis}

\begin{axis}[
height=\subfigheight,
at={(0,0)},
scale only axis,
minor x tick num=1,
xmin=0,
xmax=24,
xlabel={Time (PTU)},
ymin=0.4,
ymax=0.7,
ylabel={Price}
]
\addplot[const plot, color=plotcolor2] table[] {
0 0.46709694378946
1 0.473389001706872
2 0.464086544732755
3 0.467215921162545
4 0.47330391551742
5 0.468862484498076
6 0.456109660908683
7 0.484714343938425
8 0.488289896655922
9 0.495788878972211
10 0.491443668231034
11 0.639397694868279
12 0.629218402522711
13 0.423118660178081
14 0.420318480949411
15 0.462489293852867
16 0.441454668790988
17 0.477668463166289
18 0.499066403215792
19 0.520857294471435
20 0.649399989330777
21 0.484738346152724
22 0.462731164242489
23 0.455195037925993
24 0.455195037925993
};

\end{axis}
\end{tikzpicture}
\caption{The power program of a heat pump agent shows the power being mostly used at moments where the price is low, relieving the need to charge when the price is high; or considering the view of the grid, when a lower power consumption is required.}
\label{fig:lprh:result_hp}
\end{figure}

The power program of a heat pump agent is shown in \figref{lprh:result_hp}. This heat pump is connected directly to the congestion agent from \figref{lprh:result_ag}, hence its price profile should be identical. What is most obvious in this graph is that the high price at $\timeslot = 12$ leads to a zero power consumption, and quickly after that, the power consumption rises in order to maintain comfortable temperatures. Again, around $\timeslot = {19, 20, 21}$ the relatively high prices lead to a power consumption of zero, which was apparently feasible because of the high power consumption leading up to it.

\begin{figure}[!t]
\centering
\begin{tikzpicture}
\begin{axis}[
height=2\subfigheight,
minor x tick num=1,
xmin=0,
xmax=24,
xlabel={Time (PTU)},
ymin=-60,
ymax=60,
ylabel={Power (kW)},
ytick={-60,-40,-20,0,20,40,60},
legend style={at={(0.02,0.03)}, anchor=south west, legend cell align=left, align=left, draw=black}
]
\addplot[const plot, color=plotcolor1, style=\linestyleA] table[] {
0 19.4912500011667
1 19.5590000424167
2 17.4974999565
3 20.4934166335833
4 24.2126667005
5 25.6141666701667
6 26.677416715
7 33.9549166594167
8 36.7142498469167
9 37.3455000826667
10 39.8466666255833
11 46.2954167630833
12 40.2452500318333
13 25.0556666971667
14 25.451249934
15 31.3544167654167
16 31.6956666503333
17 38.9931666150833
18 44.7186665418333
19 49.2281666296667
20 48.2737500468333
21 39.0305000360833
22 27.69200004025
23 18.82725003775
24 18.82725003775
};
\addlegendentry{Loads}

\addplot[const plot, color=plotcolor2, style=\linestyleB] table[] {
0 -0.206083333333333
1 -0.206916666666667
2 -0.206666666666667
3 -0.206833333333333
4 -0.3695
5 -3.11625
6 -12.4484166666667
7 -24.5344166666667
8 -38.43025
9 -48.1036666666667
10 -55.9355833333333
11 -55.41175
12 -59.8324166666667
13 -56.0895
14 -50.9703333333333
15 -33.3451666666667
16 -19.9138333333333
17 -7.13208333333333
18 -1.0315
19 -0.480916666666667
20 -0.599416666666667
21 -0.445666666666667
22 -0.3445
23 -0.222666666666667
24 -0.222666666666667
};
\addlegendentry{PVs}

\addplot[const plot, color=plotcolor3, style=\linestyleC] table[] {
0 6.73854591191862
1 5.517420567187
2 7.42336750934383
3 6.92177497024212
4 5.72206951466928
5 6.67074080485696
6 9.51616530596194
7 3.48754329346652
8 2.58979737949674
9 0.893109480803156
10 2.69399349178152
11 0.797276924912655
12 5.1039239188973
13 14.8797191391483
14 16.0102172744554
15 7.8666632654862
16 11.0472571935984
17 5.71326704142715
18 0.785035184716656
19 0.676911392303156
20 0.763088607696844
21 3.17935956178017
22 7.31295275776181
23 8.5076686336332
24 8.5076686336332
};
\addlegendentry{Heat pumps}

\addplot[const plot, color=plotcolor4, style=\linestyleD] table[] {
0 0
11 -6.66942792946449
12 -2.16747774477656
13 0
18 2.18684545211142
19 0
20 -0.629941694930434
21 0
24 0
};
\addlegendentry{Batteries}

\end{axis}
\end{tikzpicture}
\caption{The total power consumption per device type show that the batteries are used far less than the heat pumps, since they have a lower efficiency. The PV panels did not have to be curtailed in this run.}
\label{fig:lprh:total_power}
\end{figure}
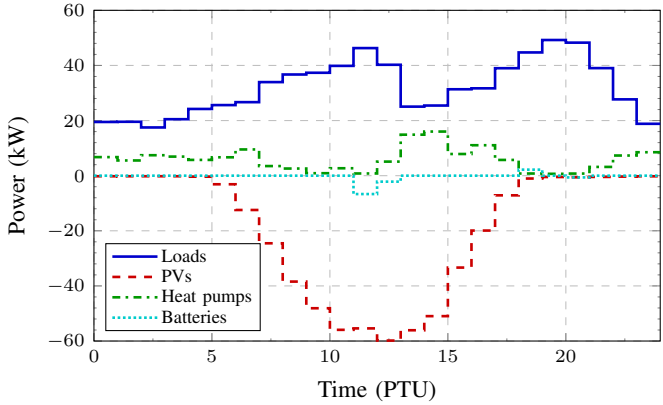

Finally, \figref{lprh:total_power} shows the total power programs of all devices of the four classes in this experiment summed up. In this figure, the power profiles of the loads and the PVs are the direct result of the chosen profiles, and are the input for problem~\eqref{eq:lprh:optim}. We can see that for the majority of the experiment, all used flexibility is from the heat pumps with the high efficiency. Only at times of the congestion will the less efficient battery agents be used.

\subsection{Comparison with Central Solvers}
\label{sec:lprh:comparison}

\begin{figure}[!t]
\centering
\begin{tikzpicture}
\begin{axis}[
height=0.25\figurewidth,
unbounded coords=jump,
minor y tick num=0,
xmin=-.5,
xmax=12,
xlabel={Power loss (kW)},
ymin=0.5,
ymax=3.5,
ytick={1,2,3},
yticklabels={{PICS},{RHCS},{LP-RH}},
ymajorgrids=false,
]

\addplot [color=black, dashed, thin] table[]{
5.12189898580723 1
7.4985469739394 1
nan nan
0 1
1.20662262459969 1
nan nan
5.33226185912071 2
7.77041825196323 2
nan nan
0 2
1.25444427403726 2
nan nan
5.57661536657665 3
8.54536041597613 3
nan nan
0 3
1.91747161660203 3
};

\addplot [color=black, thin] table[]{
7.4985469739394 0.8875
7.4985469739394 1.1125
nan nan
7.77041825196323 1.8875
7.77041825196323 2.1125
nan nan
8.54536041597613 2.8875
8.54536041597613 3.1125
nan nan
0 0.8875
0 1.1125
nan nan
0 1.8875
0 2.1125
nan nan
0 2.8875
0 3.1125
};

\addplot [color=blue, thin] table[]{
1.20662262459969 0.775
5.12189898580723 0.775
5.12189898580723 1.225
1.20662262459969 1.225
1.20662262459969 0.775
nan nan
1.25444427403726 1.775
5.33226185912071 1.775
5.33226185912071 2.225
1.25444427403726 2.225
1.25444427403726 1.775
nan nan
1.91747161660203 2.775
5.57661536657665 2.775
5.57661536657665 3.225
1.91747161660203 3.225
1.91747161660203 2.775
};

\addplot [color=red, thin] table[]{
3.30589146880462 0.775
3.30589146880462 1.225
nan nan
3.38618862107541 1.775
3.38618862107541 2.225
nan nan
3.7706366372972 2.775
3.7706366372972 3.225
};

\addplot [only marks, color=red, mark=+, thin] table[]{
9.62501651700827 1
10.3751897749075 1
9.62501651700827 2
10.8553586854899 2
9.62501651700825 3
10.1429300243827 3
11.1368502537557 3
};

\end{axis}
\end{tikzpicture}
\caption{Compared with centralized optimization solvers, the LP-RH algorithm performs equally well. In this box plot the median is shown as a line, the boxes indicate the 25\% and 75\% percentiles, and the tails are capped at a maximum length of the box width; outliers are drawn separately.}
\label{fig:lprh:solvers_compared}
\end{figure}
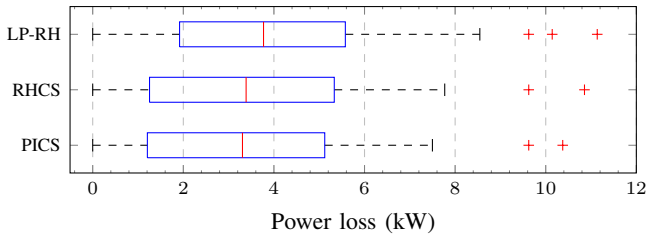

In a validation experiment 105 random problem instances (15 permutations for seven months) were created, and solved by the three different algorithms. In \figref{lprh:solvers_compared} the results are shown for the \emph{feasible} instances. A solution is considered feasible if all three solvers were able to find a solution that satisfies all constraints. The LP-RH algorithm did not find a correct solution for 26 problem instances, 11 of which were found to be overconstrained according to RHCS. For the feasible instances the LP-RH algorithm found solutions that were not significantly worse than the PICS, and in 21 instances found a solution with the exact same cost. In 27 instances LP-RH found a solution that was equally well as the RHCS or even slightly better---this seems to contradict the initial statement that RHCS acts as a lower bound for LP-RH; however, this is due to the way both solvers deal with uncertainty. The LP-RH algorithm allows the congestion agents to violate power constraints for future PTUs, as long as eventually they are not actually congested due to imperfect forecasts according to \eqref{eq:lprh:forecast}. The RHCS algorithm does not allow power constraints to be violated at any point in the future, and hence might react too strict when a future congestion is mistakenly predicted.

In the set of \emph{infeasible} problems, i.e. problems that do not have a valid solution satisfying all constraints, LP-RH does find solutions that minimize the power loss. However, since these solutions either temporarily overload congestion agents (where $\power_agen > \congestionThreshold_\agent$), or does not match the target profile $\target$ exactly, they are not a fair comparison, since they do not strictly solve \eqref{eq:lprh:optim}. In a separate run the problems were relaxed, by increasing the power rating of the congestion agents. This resulted in the LP-RH not being able to find a feasible solution in only 14 problem instances, but the results were otherwise very similar to the ones reported here in \figref{lprh:solvers_compared}.

\section{Conclusions}
We have introduced an algorithm for self-organizing smart grids, by solving the economic dispatch problem using a decentralized market based approach with local pricing. The LP-RH algorithm using a hierarchical approach was shown to be able to solve the problem using a fairly simple interaction scheme in which pricing information is sent down the hierarchy tree, and planned or forecasted power programs are sent back up. Using a gradient descent approach, the market operator is capable of tuning the pricing to find feasible solutions to minimize the power losses in the grid.

In our experiments we found that in 20\% of the problems LP-RH did not perform any worse than a perfect-information centralized solver. In the other 80\% our algorithm did not perform significantly worse. The benefit of LP-RH over a centralized solver are in the robustness and scalability of the solution, as well as in the preserved privacy of the end-consumers.

In the implementation of the response of the storage agent, we intentionally did not choose to respond with an optimal power program given the price signal. Particularly, when a high price is expected in the future, the agent will not ``proactively'' charge to avoid having to charge later, or vice versa. This behavior could be implemented at the agent quite easily using a dynamic programming approach, but it would lead to very extreme behavior, e.g. very binary behavior of charging or not-charging at full capacity even for small price differences. This binary behavior is hard to deal with in the rest of the hierarchical tree, and does not lead to any problems per se, but might be improved upon in a future continuation of this work.

Other variations of the problem may include other device types, for instance time-shiftable devices such as washing machines or dishwashers. Also, using electric vehicles (EV) as an additional type of agent, providing energy flexibility is a very interesting extension, which will undoubtedly lead to other complications because of their high power ratings. Finally, an integration with a real time balancing algorithm such as~\cite{Kok2019} would be very fruitful to complete the needs of the future smart grid.

\bibliographystyle{IEEEtran}

\end{document}